\documentclass[twocolumn,showpacs,superscriptaddress,amsmath,amssymb,aps,pra]{revtex4-2}
\usepackage{graphicx}
\usepackage{CJKutf8}
\usepackage{dcolumn}
\usepackage{amsmath,bm}
\usepackage{epstopdf}
\usepackage[mathlines]{lineno}
\usepackage{color}
\usepackage[colorlinks=true,linkcolor=blue,citecolor=magenta,urlcolor=cyan]{hyperref}

\usepackage{tikz,xcolor,hyperref}

\definecolor{lime}{HTML}{A6CE39}
\DeclareRobustCommand{\orcidicon}{%
	\begin{tikzpicture}
	\draw[lime, fill=lime] (0,0) 
	circle [radius=0.16] 
	node[white] {{\fontfamily{qag}\selectfont \tiny ID}};
	\draw[white, fill=white] (-0.068,0.105) 
	circle [radius=0.007];
	\end{tikzpicture}
	\hspace{-2mm}
}

\foreach \x in {A, ..., Z}{%
	\expandafter\xdef\csname orcid\x\endcsname{\noexpand\href{https://orcid.org/\csname orcidauthor\x\endcsname}{\noexpand\orcidicon}}
}

\begin{document}
\title{Low-energy hole subband dispersions in a cylindrical Ge nanowire: the effects of the nanowire growth direction}
\author{Rui\! Li~(\begin{CJK}{UTF8}{gbsn}李睿\end{CJK})\orcidA{}}
\email{ruili@ysu.edu.cn}
\affiliation{Key Laboratory for Microstructural Material Physics of Hebei Province, School of Science, Yanshan University, Qinhuangdao 066004, China}

\author{Zi-Qiang\! Li~(\begin{CJK}{UTF8}{gbsn}李子强\end{CJK})}
\affiliation{Key Laboratory for Microstructural Material Physics of Hebei Province, School of Science, Yanshan University, Qinhuangdao 066004, China}

\begin{abstract}

We examine the validity of the spherical approximation $\gamma_{s}=(2\gamma_{2}+3\gamma_{3})/5$ in the Luttinger-Kohn Hamiltonian in calculating the subband dispersions of the hole gas. We calculate the realistic hole subband dispersions (without the spherical approximation) in  a cylindrical Ge nanowire by using quasi-degenerate perturbation theory. The realistic low-energy hole subband dispersions have a double-well anticrossing structure, that consists with the spherical approximation prediction. However, the realistic subband dispersions are also nanowire growth direction dependent. When the nanowire growth direction is restricted in the (100) crystal plane, the detailed growth direction dependences of the subband parameters are given. We find the spherical approximation is good approximation, it can nicely reproduce the real result in some special growth directions. 
\end{abstract}
\date{May 10, 2023}
\maketitle

\section{Introduction}

Holes in the valence band of semiconductors can have distinct properties in comparison with the electrons in the conduction band~\cite{winkler2003spin,PhysRevB.62.4245}. The top of the valence band of semiconductors is four-fold degenerate, such that the minimal effective mass model of holes in bulk semiconductors is a four-band Luttinger-Kohn Hamiltonian~\cite{PhysRev.97.869,PhysRev.102.1030}. The four-band effective mass model brings about considerable calculations in the theoretical investigations, such that a series of approximations are usually made~\cite{winkler2003spin,PhysRevB.95.075305,RASHBA1988175}. The simplest approximation is the spherical approximation~\cite{PhysRev.102.1030,PhysRevB.40.8500,RASHBA1988175}, which is expected to be valid when the difference between the Luttinger parameters $\gamma_{2}$ and $\gamma_{3}$ is small,  i,e, $\gamma_{3}-\gamma_{2}\ll\gamma_{2},\gamma_{3}$.

Quasi-one-dimensional (1D) hole gas achieved in a Ge nanowire is of current interest~\cite{Lu10046,PhysRevB.93.121408,PhysRevResearch.3.013081,Froning:2021aa,Wang:2022tm,PhysRevB.87.161305,PhysRevB.105.075308,PhysRevB.106.235408}. The experiments mainly used two kinds of nanowires, i.e., the Ge/Si core/shell nanowire with a circular cross-section~\cite{Lu10046,PhysRevResearch.3.013081,Froning:2021aa} and the Ge hut wire with a triangle cross-section~\cite{Watzinger:2018aa,Gao2020AM,Wang:2022tm}. Very strong hole spin-orbit coupling has been reported in these nanowires. Measurement of the antilocalization of Coulomb blockade suggested a spin-orbit length of $20$ nm~\cite{PhysRevLett.112.216806}, while the electric-dipole spin resonance in quantum dots suggested a much shorter spin-orbit length of $1.5\sim3$ nm~\cite{Froning:2021aa,Wang:2022tm}.  Note that strong 1D hole spin-orbit coupling has potential applications in both spin quantum computing~\cite{nadj2010spin,trif2008spin,PhysRevB.88.241405,RL2013,RL2018c,RL2020,Scappucci:2021vk,PhysRevB.106.195414} and the searching of Majorana fermions~\cite{PhysRevLett.105.077001,PhysRevLett.105.177002,PhysRevB.90.195421}.

Previous theoretical studies based on the Luttinger-Kohn Hamiltonian in the spherical approximation have unveiled an interesting low-energy subband structure of the hole gas, i.e., two mutually displaced parabolic curves with an anticrossing at $k_{z}R=0$~\cite{PhysRevB.84.195314,RL2021,RL2022a}. Note that each dispersion curve is two-fold degenerate, i.e., spin degeneracy. When a strong electric field~\cite{PhysRevB.84.195314,PhysRevB.97.235422,PhysRevLett.119.126401} or a strong magnetic field~\cite{RL2022a,RL2023a,RL2023b} is used to lift this spin degeneracy, a strong linear in momentum hole spin- or `spin'-orbit coupling is achievable. Here, we address the question whether the Luttinger-Kohn Hamiltonian in the spherical approximation has reasonably given rise to the hole subband dispersions.  By using quasi-degenerate perturbation theory, we calculate the realistic hole subband dispersions in a cylindrical Ge nanowire with various growth directions. We find that the spherical approximation $\gamma_{s}=(2\gamma_{2}+3\gamma_{3})/5$ is a good approximation, it nicely reproduces the real subband dispersions in some special growth directions.  Also, the realistic hole subband dispersions are growth direction dependent, and the subband parameters as a function of the growth direction are calculated.

\section{Model of the 1D hole gas}

We study a quasi-1D hole gas confined in a cylindrical Ge nanowire. The relevant experimental setup may use a Ge/Si core/shell nanowire structure~\cite{PhysRevResearch.3.013081,Froning:2021aa}. Due to the large valence band offset at the Ge/Si interface, the transverse confining potential of the hole gas at the Ge core is well approximated by a hard-wall potential. In the framework of the effective mass approximation, the kinetic energy of the hole gas is described by the Luttinger-Kohn Hamiltonian. When the coordinate axes $k_{x,y,z}$ are along the cubic axes of the crystal, the Luttinger-Kohn Hamiltonian reads~\cite{PhysRev.97.869,PhysRev.102.1030,PhysRevB.79.155323,PhysRevB.105.075308} (in unit of $\hbar^{2}/2m$)
\begin{eqnarray}
H_{\rm LK}&=&(\gamma_{1}+\frac{5}{2}\gamma_{2})k^{2}-2\gamma_{2}({\bf k}\cdot{\bf J})^{2}\nonumber\\
&&-4(\gamma_{3}-\gamma_{2})\big(\{k_{x},k_{y}\}\{J_{x},J_{y}\}+{\rm c.p.}\big),\label{eq_LK1}
\end{eqnarray}
where $m$ is the free electron mass, $\gamma_{1}=13.35$, $\gamma_{2}=4.25$ and $\gamma_{3}=5.69$ are Luttinger parameters for semiconductor Ge~\cite{PhysRevB.4.3460}, ${\bf J}=(J_{x},J_{y},J_{z})$ is a spin-3/2 vector operator, ${\rm c.p.}$ denotes cyclic permutations, and $\{A,B\}=(AB+BA)/2$. 

\begin{figure}
\includegraphics{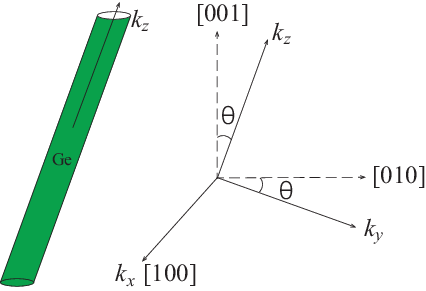}
\caption{\label{fig_coordinate}A cylindrical Ge nanowire with growth direction $\theta$ is under investigation. The $k_{x}$ axis is fixed along the crystal [100] direction, the angle between the $k_{z}$ axis and the [001] axis is denoted by $\theta$. The nanowire axis is defined as the $k_{z}$ axis.}
\end{figure}
When the coordinate axes $k_{x,y,z}$ are not along the cubic axes of the crystal, we need to rotate the Luttinger-Kohn Hamiltonian correspondingly. Here, we focus on the special case, where the $k_{x}$ axis is fixed along the [100] direction while the $k_{y}$-$k_{z}$ plane can be rotated around the $k_{x}$ axis clockwisely (see Fig.~\ref{fig_coordinate}). The rotation angle is denoted by $\theta$, i.e., the angle between the $k_{z}$ axis and the [001] crystal axis. The Luttinger-Kohn Hamiltonian in this new coordinate system reads~\cite{PhysRevB.105.L161301} (in unit of $\hbar^{2}/2m$)
\begin{eqnarray}
H_{\rm LK}&=&(\gamma_{1}+\frac{5}{2}\gamma_{2})k^{2}-2\gamma_{2}({\bf k}\cdot{\bf J})^{2}\nonumber\\
&&-4(\gamma_{3}-\gamma_{2})(\{k_{x},k_{y}\}\{J_{x},J_{y}\}+\{k_{z},k_{x}\}\{J_{z},J_{x}\})\nonumber\\
&&-(\gamma_{3}-\gamma_{2})\big[(J^{2}_{z}-J^{2}_{y})\sin2\theta+2\{J_{y},J_{z}\}\cos2\theta\big]\nonumber\\
&&\times\left(k^{2}_{z}\sin2\theta-k^{2}_{y}\sin2\theta+2k_{y}k_{z}\cos2\theta\right).\label{eq_LK2}
\end{eqnarray}
For the special angle $\theta=0$, Eq.~(\ref{eq_LK2}) can be reduced to Eq.~(\ref{eq_LK1}), just as desired.

In the following, we study the subband quantization of the hole gas for various nanowire growth directions. We choose the $k_{z}$ axis along the nanowire growth direction, e.g., $\theta=0^{\circ}$ represents the [001] growth direction and $\theta=45^{\circ}$ represents the [011] growth direction (see Fig.~\ref{fig_coordinate}). The hole Hamiltonian under investigation reads
\begin{equation}
H=H_{\rm LK}+V(r),\label{eq_H}
\end{equation}
where $V(r)$ is the transverse ($xy$ plane) confining potential
\begin{equation}
V(r)=\left\{\begin{array}{cc}0,~&~r<R,\\
\infty,~&~r>R,\end{array}\right.\label{Eq_potential}
\end{equation}
with $R$ being the radius of the Ge nanowire. Note that in our following calculations we have set $R=10$ nm, a typical and experimentally achievable nanowire radius~\cite{PhysRevLett.101.186802,PhysRevLett.112.216806}.

\section{The zeroth-order result}

Our strategy of obtaining the 1D hole subband dispersions in a general nanowire growth direction governed by Hamiltonian~(\ref{eq_H}) is to use the perturbation method~\cite{landau1965quantum}. The difference of the Luttinger parameters $\gamma_{3}-\gamma_{2}=1.44$ is relatively small in comparison with the other Luttinger parameters $\gamma_{1}=13.35$ and $\gamma_{2}=4.25$~\cite{PhysRevB.4.3460} in Eq.~(\ref{eq_LK2}), such that we treat $\gamma_{3}-\gamma_{2}$ as a perturbation parameter. The exactly solvable zeroth-order Hamiltonian reads~\cite{PhysRevB.42.3690,sweeny1988hole}
\begin{equation}
H_{0}=(\gamma_{1}+\frac{5}{2}\gamma_{2})k^{2}-2\gamma_{2}({\bf k}\cdot{\bf J})^{2}+V(r).\label{eq_H0}
\end{equation}
In the polar coordinate system where $x=r\cos\varphi$ and $y=r\sin\varphi$, the exact eigenvalues and the corresponding eigenfunctions of $H_{0}$ can be obtained with the help of both the conservation of the total angular momentum $F_{z}=J_{z}-i\partial_{\varphi}$ and the hard-wall boundary condition~\cite{PhysRevB.42.3690,sweeny1988hole,RL2021}.

\begin{figure}
\includegraphics{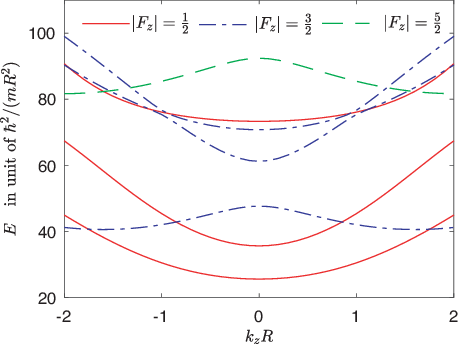} 
\caption{\label{fig_subbandH0}The hole subband dispersions calculated using the zeroth-order Hamiltonian $H_{0}$ (\ref{eq_H0}). Due to the conservation of the total angular momentum $F_{z}=J_{z}-i\partial_{\varphi}$, the subband dispersions can be classified by $|F_{z}|$. Note that each curve is two-fold degenerate and the plot is independent of the nanowire radius $R$.}
\end{figure}

Note that in the zeroth-order Hamiltonian (\ref{eq_H0}), there is also a Luttinger-Kohn Hamiltonian in the spherical approximation $\gamma_{s}=\gamma_{2}$.  We also emphasize that the usually used spherical approximation in the literature is $\gamma_{s}=(2\gamma_{2}+3\gamma_{3})/5$~\cite{PhysRevB.84.195314}. The zeroth-order hole subband dispersions are shown in Fig.~\ref{fig_subbandH0}. Surprisingly, the lowest two subband dispersions given by $|F_{z}|=1/2$ do not show a double-well anticrossing structure, i.e., two mutually displaced parabolic curves with an anticrossing at $k_{z}R=0$~\cite{PhysRevB.84.195314,RL2021}, that is obtainable when $\gamma_{2}$ in Eq.~(\ref{eq_H0}) is replaced by $\gamma_{s}=(2\gamma_{2}+3\gamma_{3})/5$. This discrepancy indicates that the shape of the low-energy hole subband dispersions is sensitive to the value of $\gamma_{s}$ in the spherical approximation. Therefore, it is an interesting question that which choice of $\gamma_{s}$ gives the reasonable hole subband dispersions. Our perturbation calculations in the following will answer this question. 

The eigenfunctions of $H_{0}$ are also important in the following perturbation calculations. We note that each energy level of $H_{0}$ is two-fold degenerate, i.e., each subband curve in Fig.~\ref{fig_subbandH0} is two-fold (spin) degenerate. Also, the degenerate partner of a given eigenfunction can be found from symmetry analysis~\cite{RASHBA1988175,RL2021}. The detailed expressions of these two degenerate eigenfunctions can be found elsewhere~\cite{RL2021}.

\section{The perturbation result}

Since both the zeroth-order eigenvalues and the corresponding eigenfunctions are obtainable, we can solve the eigenvalues of Hamiltonian (\ref{eq_H}) using quasi-degenerate perturbation theory~\cite{landau1965quantum}. In order to guarantee enough calculation accuracy, we have chosen an eight dimensional quasi-degenerate Hilbert subspace in the following calculations. The detailed basis states are chosen as follows: six states from the lowest three subbands of $|F_{z}|=1/2$ and two states from the lowest subband of $|F_{z}|=3/2$ (for details see Appendix~\ref{appendix}). The Hamiltonian (\ref{eq_H}) can be written as a $8\times8$ matrix in this Hilbert subspace.

\begin{figure}
\includegraphics{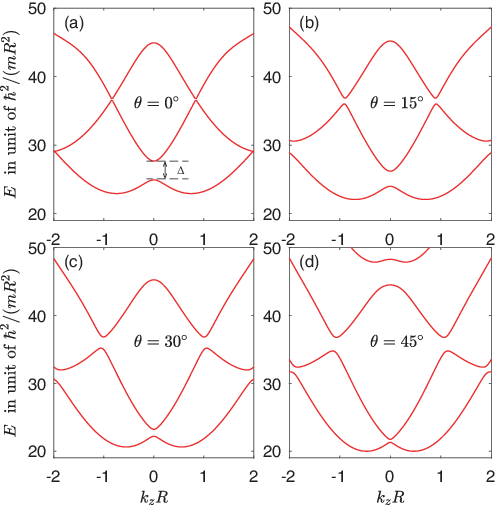}
\caption{\label{fig_subbandH}The hole subband dispersions calculated using quasi-degenerate perturbation theory. The Hamiltonian (\ref{eq_H}) is written as a $8\times8$ matrix in the quasi-degenerate Hilbert subspace.  The results of nanowire growth directions $\theta=0^{\circ}$ (a), $\theta=15^{\circ}$ (b), $\theta=30^{\circ}$ (c), and $\theta=45^{\circ}$ (d). Note that the energy unit $\hbar^{2}/(mR^{2})\approx0.763$ meV for radius $R=10$ nm, and each dispersion curve is still two-fold degenerate.}
\end{figure}

Diagonalizing the $8\times8$ matrix of $H$, we obtain four subband dispersions, each of which is still two-fold degenerate. We show in Fig.~\ref{fig_subbandH} the lowest three hole subband dispersions in various nanowire growth directions. Interestingly, our quasi-degenerate perturbation calculation nicely gives rise to a double-well anticrossing subband structure, which is missing in the zeroth-order result (see Fig.~\ref{fig_subbandH0}). These results are qualitatively similar to that obtained using the spherical approximation where  $\gamma_{2}$ in Eq.~(\ref{eq_H0}) is replaced by $\gamma_{s}=(2\gamma_{2}+3\gamma_{3})/5$~\cite{PhysRevB.84.195314,RL2021}. Despite this similarity, the subband parameters $m^{*}_{h}$, $\alpha$, and $\Delta$ [see Eq.~\ref{eq_Hamiltoniankz2} below] also depend on the nanowire growth direction $\theta$.

\begin{figure}
\includegraphics{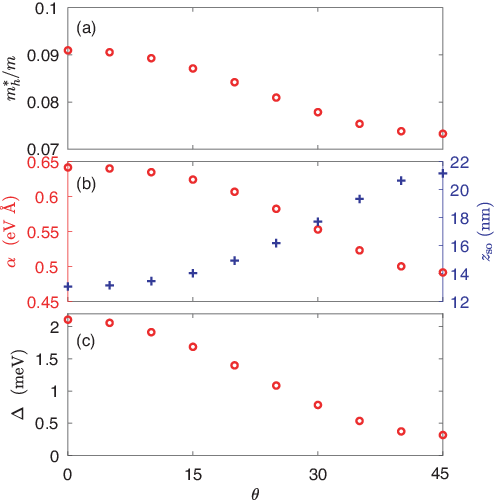}
\caption{\label{fig_parameters}The subband parameters $m^{*}_{h}$, $\alpha$, and $\Delta$ as a function of the nanowire growth direction $\theta$. The results of the effective mass $m^{*}_{h}$ (a), both the `spin'-orbit coupling strength $\alpha$ and the `spin'-orbit length $z_{\rm so}=\hbar^{2}/(m^{*}_{h}\alpha)$ (b), and the energy gap $\Delta$ (c).}
\end{figure}

The lowest two hole subband dispersions, i.e., two mutually displaced parabolic curves with an anticrossing at $k_{z}R=0$, can be approximately described by the following effective Hamiltonian~\cite{PhysRevB.84.195314,RL2023b}
\begin{equation}
H^{\rm ef}\approx\frac{\hbar^{2}k^{2}_{z}}{2m^{*}_{h}}+\alpha\,k_{z}\tau^{x}s^{x}+\frac{\Delta}{2}\tau^{z},\label{eq_Hamiltoniankz2}
\end{equation}
where $m^{*}_{h}$ is the effective hole mass, $\alpha$ is the `spin'-orbit coupling strength, $\Delta$ is the energy gap at the anticrossing site $k_{z}R=0$, $\boldsymbol \tau$ is the `spin' (pseudo spin) operator defined in the Hilbert subspace spanned by the lowest two orbital states at $k_{z}R=0$, and ${\bf s}$ is the real hole spin operator.

The subband parameters $m^{*}_{h}$, $\alpha$, and $\Delta$ as a function of the nanowire growth direction $\theta$ are shown in Figs.~\ref{fig_parameters}(a), (b), and (c), respectively. Here, $m^{*}_{h}$, $\alpha$, and $\Delta$ are obtained by a band fitting. Note that these parameters have very similar $\theta$ dependence, e.g., they are largest in growth direction $\theta=0$, i.e., the [001] growth direction, they are smallest in growth direction $\theta=\pi/4$, i.e., the [011] growth direction. Also, the `spin'-orbit length $z_{\rm so}=\hbar^{2}/(m^{*}_{h}\alpha)$ as a function of $\theta$ can be found in Fig.~\ref{fig_parameters}(b). For a nanowire with radius $R=10$ nm, the `spin'-orbit length $z_{\rm so}$ changes from 12 to 22 nms with the increase of $\theta$, and smaller $R$ will lead to smaller $z_{\rm so}$. 

\begin{figure}
\includegraphics{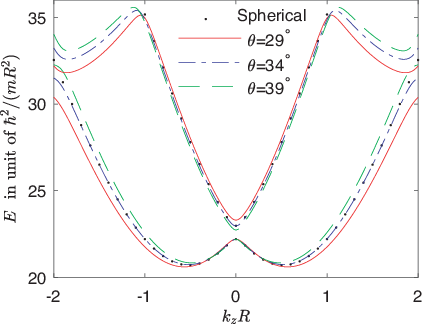}
\caption{\label{fig_comparsion}The lowest two hole subband dispersions near the growth direction $\theta=34^{\circ}$. The result (dotted lines) of the spherical approximation $\gamma_{s}=(2\gamma_{2}+3\gamma_{3})/5$ nicely coincides with that of the nanowire growth direction $\theta=34^{\circ}$.}
\end{figure}

Let us discuss the validity of using the Luttinger-Kohn Hamiltonian in the spherical approximation in calculating the hole subband dispersions. First, the spherical approximation, i.e.,  $\gamma_{2}$ in Eq.~(\ref{eq_H0}) is replaced by $\gamma_{s}=(2\gamma_{2}+3\gamma_{3})/5$~\cite{PhysRevB.84.195314}, indeed qualitatively reproduces the low-energy hole subband structure in the nanowire, i.e., the double-well anticrossing structure. Note that the simple choice $\gamma_{s}=\gamma_{2}$ cannot give rise to the reasonable hole subband structure (see Fig.~\ref{fig_subbandH0}). Second, the choice $\gamma_{s}=(2\gamma_{2}+3\gamma_{3})/5$ nicely reproduces the hole subband dispersions~\cite{PhysRevB.84.195314,RL2022a} in some special nanowire growth directions, e.g., $\theta\approx34^{\circ}$ in our case (see Fig.~\ref{fig_comparsion}). Note that an overall constant energy is added or removed to each $\theta$ line in order to let all the dispersions in Fig.~\ref{fig_comparsion} have the same energy at $k_{z}R=0$. 

\section{The effects of static strain}

\begin{figure}
\includegraphics{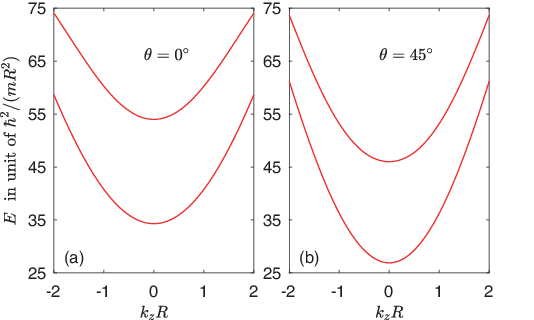}
\caption{\label{fig_strain}The lowest two hole subband dispersions in a Ge/Si core/shell nanowire in the presence of static strain. The results in the [001] growth direction (a) and [011] growth direction (b). The relative shell thickness is chosen as $(R_{s}-R)/R=0.2$, where $R_{s}$ is the radius of the Si shell.}
\end{figure}

A Ge/Si core/shell nanowire is usually used in experiments for hosting the 1D hole gas. When lattice-mismatched heterostructures are prepared using the pseudomorphic growth technique, static strain will be induced in the growth layer~\cite{sun2009strain}. Here, we study the case of a strained Ge core, where the lattice constant of the Ge core tends to match the lattice constant of the Si shell. The strain effects on the hole states are described by the Bir-Pikus Hamiltonian~\cite{bir1974symmetry}
\begin{equation}
H_{\rm BP}=b\sum_{i=x,y,z}\varepsilon_{ii}J^{2}_{i}+\frac{2d}{\sqrt{3}}(\varepsilon_{xy}\{J_{x},J_{y}\}+{\rm c.p.}),
\end{equation}
where $b\approx-2.5$ eV and $d\approx-5.0$ eV are material parameters of Ge, and $x,y$ and $z$ are along the cubic axes of the crystal. When the coordinate axes are not along the lattice cubic axes, e.g., for a general $\theta$ growth direction shown in Fig.~\ref{fig_coordinate}, we should first obtain the corresponding Bir-Pikus Hamiltonian using coordinate transformation.  The elements of the strain tensor read $\varepsilon_{xx}=\varepsilon_{yy}=\varepsilon_{\perp}$, $\varepsilon_{zz}=\varepsilon_{\parallel}$, and $\varepsilon_{xy}=\varepsilon_{xz}=\varepsilon_{yz}=0$. The Bir-Pikus Hamiltonian for $\theta$ growth direction can be obtained following Ref.~\cite{PhysRevB.90.115419}
\begin{eqnarray}
H_{\rm BP}&=&-b(\varepsilon_{\perp}-\varepsilon_{\parallel})J^{2}_{z}+\frac{1}{2}\left(b-\frac{d}{\sqrt{3}}\right)(\varepsilon_{\perp}-\varepsilon_{\parallel})\times\nonumber\\
&&\left[(J^{2}_{z}-J^{2}_{y})\sin^{2}2\theta+\{J_{y},J_{z}\}\sin4\theta\right].
\end{eqnarray}
Note that the magnitude of the second term is much smaller than that of the first term, and the second term vanishes when the spherical approximation $d=\sqrt{3}b$ is used. For a core/shell nanowire with a Ge core  $R=10$ nm and a Si shell $R_{s}=12$ nm, the non-zero elements of the strain tensor read $\varepsilon_{\perp}\approx-0.0037$ and $\varepsilon_{\parallel}\approx-0.0144$~\cite{PhysRevB.90.115419}. In the quasi-degenerate Hilbert subspace, $H_{\rm BP}$ is written as a $8\times8$ matrix, we can diagonalize the total Hamiltonian $H+H_{\rm BP}$ to obtain the low-energy hole subband dispersions. 

The presence of the strain destroys the double-well anticrossing structure (see Fig.~\ref{fig_subbandH}) in the low-energy subband dispersions. We show in Figs.~\ref{fig_strain}(a) and (b) the lowest two subband dispersions in the [001] and [011] nanowire growth directions, respectively. The lowest hole subband dispersion now is a single parabolic curve, and the lowest two subband dispersions are separated significantly by the static strain. If we still use Eq.~(\ref{eq_Hamiltoniankz2}) to model these two lowest subband dispersions, the `spin'-orbit coupling energy now is much less than the Zeeman energy, i.e., $m^{*}_{h}\alpha^{2}\ll\Delta/2$.

\section{The accuracy of the perturbation calculation}

\begin{figure}
\includegraphics{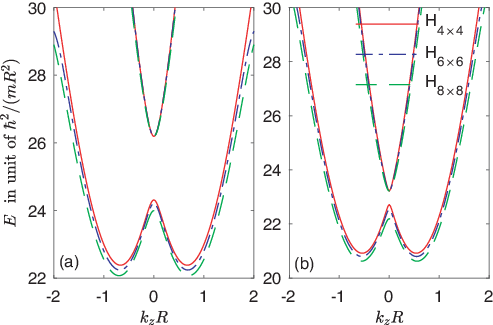}
\caption{\label{fig_stable}The lowest two hole subband dispersions calculated using different dimensions of the Hilbert subspace. The results of nanowire growth directions $\theta=15^{\circ}$ (a) and $\theta=30^{\circ}$ (b).}
\end{figure}

In our perturbation calculation, we have chosen an eight dimensional quasi-degenerate Hilbert subpace. The perturbation calculation gives rise to the well-known low-energy double-well anticrossing band structure, that consists with both the numerical and the analytical calculations in the literature~\cite{PhysRevB.84.195314,PhysRevB.97.235422,RL2021,RL2022a,doi:10.1063/1.4929412,doi:10.1063/1.4972987}. Here, we study the accuracy (or the stability) of our perturbation calculation. The stability of our calculation can be demonstrated by analyzing the results of increasing the dimension of the quasi-degenerate Hilbert subspace.

We show in Fig.~\ref{fig_stable} the perturbation results using different dimensions of the quasi-degenerate Hilbert subspace. Figures~\ref{fig_stable}(a) and (b) show the results of the nanowire growth directions $\theta=15^{\circ}$ and $\theta=30^{\circ}$, respectively.  For each nanowire growth direction, the results of using 4, 6, and 8 dimensional Hilbert subspace are given. Increasing the dimension of the Hilbert subspace only gives very small rectification to the subband dispersions, and it does not change the double-well anticrossing band structure. Also, choosing different dimension of the Hilbert subspace does not change the directional dependences of the subband parameters $m^{*}_{h}$, $\alpha$, and $\Delta$. For example, in the same dimension of Hilbert subspace, the band gap $\Delta$ of $\theta=15^{\circ}$ [see Fig.~\ref{fig_stable}(a)] is apparently larger than that of $\theta=30^{\circ}$ [see Fig.~\ref{fig_stable}(b)]. Therefore, our perturbation results not only qualitatively but also quantitatively reflect the growth directional dependence of the low-energy hole subband dispersions.

\section{Discussion and summary}

Although the quasi-1D hole subband dispersions are usually calculated theoretically, it is also of importance to determine them from the experimental aspect. Similar to the quasi-2D hole gas in quantum well, the quasi-1D hole subband dispersions may be directly measured using optical method~\cite{KASH1994251}.  Besides, the effective hole mass $m^{*}_{h}$ gives some indirect information on the subband dispersions, such that another useful way is to extract $m^{*}_{h}$ from experimental measurements. 

In summary, when the nanowire growth direction is restricted in the (100) crystal plane, we have calculated the growth direction dependence of the low-energy hole subband dispersions in a cylindrical Ge nanowire. The realistic low-energy hole subband dispersions indeed show a double-well anticrossing structure, i.e., two mutually displaced parabolic curves with an anticrossing at $k_{z}R=0$, that consists with the spherical approximation $\gamma_{s}=(2\gamma_{2}+3\gamma_{3})/5$ prediction. However, the low-energy subband parameters $m^{*}_{h}$, $\alpha$, and $\Delta$ are also nanowire growth direction dependent. They are largest in the [001] ($\theta=0^{\circ}$) growth direction, and they are smallest in the [011] ($\theta=45^{\circ}$) growth direction. The spherical approximation can nicely reproduce the real low-energy hole subband dispersions in some special growth directions.

\section*{Acknowledgements}
This work was supported by the National Natural Science Foundation of China Grant No.~11404020, the Project from the Department of Education of Hebei Province Grant No. QN2019057, and the Starting up Foundation from Yanshan University Grant No. BL18043.

\appendix
\section{\label{appendix}Basis states of the quasi-degenerate perturbation calculation}
The basis states of the quasi-degenerate perturbation calculation are chosen from the eigenstates of Hamiltonian $H_{0}$ (\ref{eq_H0}). The eight basis states read 
\begin{eqnarray}
|1\rangle&=&\left(\begin{array}{c}\Psi_{a1}(r)e^{-i\varphi}\\\Psi_{a2}(r)\\\Psi_{a3}(r)e^{i\varphi}\\\Psi_{a4}(r)e^{2i\varphi}\end{array}\right),~|2\rangle=\left(\begin{array}{c}\Psi^{*}_{a4}(r)e^{-2i\varphi}\\\Psi^{*}_{a3}(r)e^{-i\varphi}\\\Psi^{*}_{a2}(r)\\\Psi^{*}_{a1}(r)e^{i\varphi}\end{array}\right),\nonumber\\
|3\rangle&=&\left(\begin{array}{c}\Psi_{b1}(r)e^{-i\varphi}\\\Psi_{b2}(r)\\\Psi_{b3}(r)e^{i\varphi}\\\Psi_{b4}(r)e^{2i\varphi}\end{array}\right),~|4\rangle=\left(\begin{array}{c}\Psi^{*}_{b4}(r)e^{-2i\varphi}\\\Psi^{*}_{b3}(r)e^{-i\varphi}\\\Psi^{*}_{b2}(r)\\\Psi^{*}_{b1}(r)e^{i\varphi}\end{array}\right),\nonumber\\
|5\rangle&=&\left(\begin{array}{c}\Psi_{c1}(r)\\\Psi_{c2}(r)e^{i\varphi}\\\Psi_{c3}(r)e^{2i\varphi}\\\Psi_{c4}(r)e^{3i\varphi}\end{array}\right),~|6\rangle=\left(\begin{array}{c}\Psi^{*}_{c4}(r)e^{-3i\varphi}\\\Psi^{*}_{c3}(r)e^{-2i\varphi}\\\Psi^{*}_{c2}(r)e^{-i\varphi}\\\Psi^{*}_{c1}(r)
\end{array}\right),\nonumber\\
|7\rangle&=&\left(\begin{array}{c}\Psi_{d1}(r)e^{-i\varphi}\\\Psi_{d2}(r)\\\Psi_{d3}(r)e^{i\varphi}\\\Psi_{d4}(r)e^{2i\varphi}\end{array}\right),~|8\rangle=\left(\begin{array}{c}\Psi^{*}_{d4}(r)e^{-2i\varphi}\\\Psi^{*}_{d3}(r)e^{-i\varphi}\\\Psi^{*}_{d2}(r)\\\Psi^{*}_{d1}(r)e^{i\varphi}\end{array}\right),
\end{eqnarray}
where the expressions of $\Psi_{a,b,c,d1}(r)$, $\Psi_{a,b,c,d2}(r)$, $\Psi_{a,b,c,d3}(r)$, and $\Psi_{a,b,c,d4}(r)$ can be found in Ref.~\cite{RL2021}. States $|1\rangle$, $|3\rangle$, and $|7\rangle$ are the lowest three eigenstates of $F_{z}=1/2$, and state $|5\rangle$ is the lowest eigenstate of $F_{z}=3/2$. States $|2\rangle$, $|4\rangle$, $|6\rangle$, and $|8\rangle$ are the degenerate partners of states $|1\rangle$, $|3\rangle$, $|5\rangle$, and $|7\rangle$, respectively. 
\bibliography{Ref_Hole_spin}

\begin{thebibliography}{49}%
\makeatletter
\providecommand \@ifxundefined [1]{%
 \@ifx{#1\undefined}
}%
\providecommand \@ifnum [1]{%
 \ifnum #1\expandafter \@firstoftwo
 \else \expandafter \@secondoftwo
 \fi
}%
\providecommand \@ifx [1]{%
 \ifx #1\expandafter \@firstoftwo
 \else \expandafter \@secondoftwo
 \fi
}%
\providecommand \natexlab [1]{#1}%
\providecommand \enquote  [1]{``#1''}%
\providecommand \bibnamefont  [1]{#1}%
\providecommand \bibfnamefont [1]{#1}%
\providecommand \citenamefont [1]{#1}%
\providecommand \href@noop [0]{\@secondoftwo}%
\providecommand \href [0]{\begingroup \@sanitize@url \@href}%
\providecommand \@href[1]{\@@startlink{#1}\@@href}%
\providecommand \@@href[1]{\endgroup#1\@@endlink}%
\providecommand \@sanitize@url [0]{\catcode `\\12\catcode `\$12\catcode
  `\&12\catcode `\#12\catcode `\^12\catcode `\_12\catcode `\%12\relax}%
\providecommand \@@startlink[1]{}%
\providecommand \@@endlink[0]{}%
\providecommand \url  [0]{\begingroup\@sanitize@url \@url }%
\providecommand \@url [1]{\endgroup\@href {#1}{\urlprefix }}%
\providecommand \urlprefix  [0]{URL }%
\providecommand \Eprint [0]{\href }%
\providecommand \doibase [0]{https://doi.org/}%
\providecommand \selectlanguage [0]{\@gobble}%
\providecommand \bibinfo  [0]{\@secondoftwo}%
\providecommand \bibfield  [0]{\@secondoftwo}%
\providecommand \translation [1]{[#1]}%
\providecommand \BibitemOpen [0]{}%
\providecommand \bibitemStop [0]{}%
\providecommand \bibitemNoStop [0]{.\EOS\space}%
\providecommand \EOS [0]{\spacefactor3000\relax}%
\providecommand \BibitemShut  [1]{\csname bibitem#1\endcsname}%
\let\auto@bib@innerbib\@empty
\bibitem [{\citenamefont {Winkler}(2003)}]{winkler2003spin}%
  \BibitemOpen
  \bibfield  {author} {\bibinfo {author} {\bibfnamefont {R.}~\bibnamefont
  {Winkler}},\ }\href@noop {} {\emph {\bibinfo {title} {Spin-Orbit Coupling
  Effects in Two-Dimensional Electron and Hole Systems}}}\ (\bibinfo
  {publisher} {Springer, Berlin},\ \bibinfo {year} {2003})\BibitemShut
  {NoStop}%
\bibitem [{\citenamefont {Winkler}(2000)}]{PhysRevB.62.4245}%
  \BibitemOpen
  \bibfield  {author} {\bibinfo {author} {\bibfnamefont {R.}~\bibnamefont
  {Winkler}},\ }\bibfield  {title} {\bibinfo {title} {Rashba spin splitting in
  two-dimensional electron and hole systems},\ }\href
  {https://doi.org/10.1103/PhysRevB.62.4245} {\bibfield  {journal} {\bibinfo
  {journal} {Phys. Rev. B}\ }\textbf {\bibinfo {volume} {62}},\ \bibinfo
  {pages} {4245} (\bibinfo {year} {2000})}\BibitemShut {NoStop}%
\bibitem [{\citenamefont {Luttinger}\ and\ \citenamefont
  {Kohn}(1955)}]{PhysRev.97.869}%
  \BibitemOpen
  \bibfield  {author} {\bibinfo {author} {\bibfnamefont {J.~M.}\ \bibnamefont
  {Luttinger}}\ and\ \bibinfo {author} {\bibfnamefont {W.}~\bibnamefont
  {Kohn}},\ }\bibfield  {title} {\bibinfo {title} {Motion of electrons and
  holes in perturbed periodic fields},\ }\href
  {https://doi.org/10.1103/PhysRev.97.869} {\bibfield  {journal} {\bibinfo
  {journal} {Phys. Rev.}\ }\textbf {\bibinfo {volume} {97}},\ \bibinfo {pages}
  {869} (\bibinfo {year} {1955})}\BibitemShut {NoStop}%
\bibitem [{\citenamefont {Luttinger}(1956)}]{PhysRev.102.1030}%
  \BibitemOpen
  \bibfield  {author} {\bibinfo {author} {\bibfnamefont {J.~M.}\ \bibnamefont
  {Luttinger}},\ }\bibfield  {title} {\bibinfo {title} {Quantum theory of
  cyclotron resonance in semiconductors: General theory},\ }\href
  {https://doi.org/10.1103/PhysRev.102.1030} {\bibfield  {journal} {\bibinfo
  {journal} {Phys. Rev.}\ }\textbf {\bibinfo {volume} {102}},\ \bibinfo {pages}
  {1030} (\bibinfo {year} {1956})}\BibitemShut {NoStop}%
\bibitem [{\citenamefont {Marcellina}\ \emph {et~al.}(2017)\citenamefont
  {Marcellina}, \citenamefont {Hamilton}, \citenamefont {Winkler},\ and\
  \citenamefont {Culcer}}]{PhysRevB.95.075305}%
  \BibitemOpen
  \bibfield  {author} {\bibinfo {author} {\bibfnamefont {E.}~\bibnamefont
  {Marcellina}}, \bibinfo {author} {\bibfnamefont {A.~R.}\ \bibnamefont
  {Hamilton}}, \bibinfo {author} {\bibfnamefont {R.}~\bibnamefont {Winkler}},\
  and\ \bibinfo {author} {\bibfnamefont {D.}~\bibnamefont {Culcer}},\
  }\bibfield  {title} {\bibinfo {title} {Spin-orbit interactions in
  inversion-asymmetric two-dimensional hole systems: A variational analysis},\
  }\href {https://doi.org/10.1103/PhysRevB.95.075305} {\bibfield  {journal}
  {\bibinfo  {journal} {Phys. Rev. B}\ }\textbf {\bibinfo {volume} {95}},\
  \bibinfo {pages} {075305} (\bibinfo {year} {2017})}\BibitemShut {NoStop}%
\bibitem [{\citenamefont {Rashba}\ and\ \citenamefont
  {Sherman}(1988)}]{RASHBA1988175}%
  \BibitemOpen
  \bibfield  {author} {\bibinfo {author} {\bibfnamefont {E.}~\bibnamefont
  {Rashba}}\ and\ \bibinfo {author} {\bibfnamefont {E.}~\bibnamefont
  {Sherman}},\ }\bibfield  {title} {\bibinfo {title} {Spin-orbital band
  splitting in symmetric quantum wells},\ }\href
  {https://doi.org/https://doi.org/10.1016/0375-9601(88)90140-5} {\bibfield
  {journal} {\bibinfo  {journal} {Physics Letters A}\ }\textbf {\bibinfo
  {volume} {129}},\ \bibinfo {pages} {175} (\bibinfo {year}
  {1988})}\BibitemShut {NoStop}%
\bibitem [{\citenamefont {Xia}(1989)}]{PhysRevB.40.8500}%
  \BibitemOpen
  \bibfield  {author} {\bibinfo {author} {\bibfnamefont {J.-B.}\ \bibnamefont
  {Xia}},\ }\bibfield  {title} {\bibinfo {title} {Electronic structures of
  zero-dimensional quantum wells},\ }\href
  {https://doi.org/10.1103/PhysRevB.40.8500} {\bibfield  {journal} {\bibinfo
  {journal} {Phys. Rev. B}\ }\textbf {\bibinfo {volume} {40}},\ \bibinfo
  {pages} {8500} (\bibinfo {year} {1989})}\BibitemShut {NoStop}%
\bibitem [{\citenamefont {Lu}\ \emph {et~al.}(2005)\citenamefont {Lu},
  \citenamefont {Xiang}, \citenamefont {Timko}, \citenamefont {Wu},\ and\
  \citenamefont {Lieber}}]{Lu10046}%
  \BibitemOpen
  \bibfield  {author} {\bibinfo {author} {\bibfnamefont {W.}~\bibnamefont
  {Lu}}, \bibinfo {author} {\bibfnamefont {J.}~\bibnamefont {Xiang}}, \bibinfo
  {author} {\bibfnamefont {B.~P.}\ \bibnamefont {Timko}}, \bibinfo {author}
  {\bibfnamefont {Y.}~\bibnamefont {Wu}},\ and\ \bibinfo {author}
  {\bibfnamefont {C.~M.}\ \bibnamefont {Lieber}},\ }\bibfield  {title}
  {\bibinfo {title} {One-dimensional hole gas in germanium/silicon nanowire
  heterostructures},\ }\href {https://doi.org/10.1073/pnas.0504581102}
  {\bibfield  {journal} {\bibinfo  {journal} {Proceedings of the National
  Academy of Sciences}\ }\textbf {\bibinfo {volume} {102}},\ \bibinfo {pages}
  {10046} (\bibinfo {year} {2005})}\BibitemShut {NoStop}%
\bibitem [{\citenamefont {Brauns}\ \emph {et~al.}(2016)\citenamefont {Brauns},
  \citenamefont {Ridderbos}, \citenamefont {Li}, \citenamefont {Bakkers},\ and\
  \citenamefont {Zwanenburg}}]{PhysRevB.93.121408}%
  \BibitemOpen
  \bibfield  {author} {\bibinfo {author} {\bibfnamefont {M.}~\bibnamefont
  {Brauns}}, \bibinfo {author} {\bibfnamefont {J.}~\bibnamefont {Ridderbos}},
  \bibinfo {author} {\bibfnamefont {A.}~\bibnamefont {Li}}, \bibinfo {author}
  {\bibfnamefont {E.~P. A.~M.}\ \bibnamefont {Bakkers}},\ and\ \bibinfo
  {author} {\bibfnamefont {F.~A.}\ \bibnamefont {Zwanenburg}},\ }\bibfield
  {title} {\bibinfo {title} {{Electric-field dependent $g$-factor anisotropy in
  Ge-Si core-shell nanowire quantum dots}},\ }\href
  {https://doi.org/10.1103/PhysRevB.93.121408} {\bibfield  {journal} {\bibinfo
  {journal} {Phys. Rev. B}\ }\textbf {\bibinfo {volume} {93}},\ \bibinfo
  {pages} {121408} (\bibinfo {year} {2016})}\BibitemShut {NoStop}%
\bibitem [{\citenamefont {Froning}\ \emph
  {et~al.}(2021{\natexlab{a}})\citenamefont {Froning}, \citenamefont
  {Ran\ifmmode \check{c}\else \v{c}\fi{}i\ifmmode~\acute{c}\else \'{c}\fi{}},
  \citenamefont {Het\'enyi}, \citenamefont {Bosco}, \citenamefont {Rehmann},
  \citenamefont {Li}, \citenamefont {Bakkers}, \citenamefont {Zwanenburg},
  \citenamefont {Loss}, \citenamefont {Zumb\"uhl},\ and\ \citenamefont
  {Braakman}}]{PhysRevResearch.3.013081}%
  \BibitemOpen
  \bibfield  {author} {\bibinfo {author} {\bibfnamefont {F.~N.~M.}\
  \bibnamefont {Froning}}, \bibinfo {author} {\bibfnamefont {M.~J.}\
  \bibnamefont {Ran\ifmmode \check{c}\else \v{c}\fi{}i\ifmmode~\acute{c}\else
  \'{c}\fi{}}}, \bibinfo {author} {\bibfnamefont {B.}~\bibnamefont
  {Het\'enyi}}, \bibinfo {author} {\bibfnamefont {S.}~\bibnamefont {Bosco}},
  \bibinfo {author} {\bibfnamefont {M.~K.}\ \bibnamefont {Rehmann}}, \bibinfo
  {author} {\bibfnamefont {A.}~\bibnamefont {Li}}, \bibinfo {author}
  {\bibfnamefont {E.~P. A.~M.}\ \bibnamefont {Bakkers}}, \bibinfo {author}
  {\bibfnamefont {F.~A.}\ \bibnamefont {Zwanenburg}}, \bibinfo {author}
  {\bibfnamefont {D.}~\bibnamefont {Loss}}, \bibinfo {author} {\bibfnamefont
  {D.~M.}\ \bibnamefont {Zumb\"uhl}},\ and\ \bibinfo {author} {\bibfnamefont
  {F.~R.}\ \bibnamefont {Braakman}},\ }\bibfield  {title} {\bibinfo {title}
  {Strong spin-orbit interaction and $g$-factor renormalization of hole spins
  in {G}e/{S}i nanowire quantum dots},\ }\href
  {https://doi.org/10.1103/PhysRevResearch.3.013081} {\bibfield  {journal}
  {\bibinfo  {journal} {Phys. Rev. Research}\ }\textbf {\bibinfo {volume}
  {3}},\ \bibinfo {pages} {013081} (\bibinfo {year}
  {2021}{\natexlab{a}})}\BibitemShut {NoStop}%
\bibitem [{\citenamefont {Froning}\ \emph
  {et~al.}(2021{\natexlab{b}})\citenamefont {Froning}, \citenamefont
  {Camenzind}, \citenamefont {van~der Molen}, \citenamefont {Li}, \citenamefont
  {Bakkers}, \citenamefont {Zumb{\"u}hl},\ and\ \citenamefont
  {Braakman}}]{Froning:2021aa}%
  \BibitemOpen
  \bibfield  {author} {\bibinfo {author} {\bibfnamefont {F.~N.~M.}\
  \bibnamefont {Froning}}, \bibinfo {author} {\bibfnamefont {L.~C.}\
  \bibnamefont {Camenzind}}, \bibinfo {author} {\bibfnamefont {O.~A.~H.}\
  \bibnamefont {van~der Molen}}, \bibinfo {author} {\bibfnamefont
  {A.}~\bibnamefont {Li}}, \bibinfo {author} {\bibfnamefont {E.~P. A.~M.}\
  \bibnamefont {Bakkers}}, \bibinfo {author} {\bibfnamefont {D.~M.}\
  \bibnamefont {Zumb{\"u}hl}},\ and\ \bibinfo {author} {\bibfnamefont {F.~R.}\
  \bibnamefont {Braakman}},\ }\bibfield  {title} {\bibinfo {title} {Ultrafast
  hole spin qubit with gate-tunable spin--orbit switch functionality},\ }\href
  {https://doi.org/10.1038/s41565-020-00828-6} {\bibfield  {journal} {\bibinfo
  {journal} {Nature Nanotechnology}\ }\textbf {\bibinfo {volume} {16}},\
  \bibinfo {pages} {308} (\bibinfo {year} {2021}{\natexlab{b}})}\BibitemShut
  {NoStop}%
\bibitem [{\citenamefont {Wang}\ \emph {et~al.}(2022)\citenamefont {Wang},
  \citenamefont {Xu}, \citenamefont {Gao}, \citenamefont {Liu}, \citenamefont
  {Ma}, \citenamefont {Zhang}, \citenamefont {Wang}, \citenamefont {Cao},
  \citenamefont {Wang}, \citenamefont {Zhang}, \citenamefont {Culcer},
  \citenamefont {Hu}, \citenamefont {Jiang}, \citenamefont {Li}, \citenamefont
  {Guo},\ and\ \citenamefont {Guo}}]{Wang:2022tm}%
  \BibitemOpen
  \bibfield  {author} {\bibinfo {author} {\bibfnamefont {K.}~\bibnamefont
  {Wang}}, \bibinfo {author} {\bibfnamefont {G.}~\bibnamefont {Xu}}, \bibinfo
  {author} {\bibfnamefont {F.}~\bibnamefont {Gao}}, \bibinfo {author}
  {\bibfnamefont {H.}~\bibnamefont {Liu}}, \bibinfo {author} {\bibfnamefont
  {R.-L.}\ \bibnamefont {Ma}}, \bibinfo {author} {\bibfnamefont
  {X.}~\bibnamefont {Zhang}}, \bibinfo {author} {\bibfnamefont
  {Z.}~\bibnamefont {Wang}}, \bibinfo {author} {\bibfnamefont {G.}~\bibnamefont
  {Cao}}, \bibinfo {author} {\bibfnamefont {T.}~\bibnamefont {Wang}}, \bibinfo
  {author} {\bibfnamefont {J.-J.}\ \bibnamefont {Zhang}}, \bibinfo {author}
  {\bibfnamefont {D.}~\bibnamefont {Culcer}}, \bibinfo {author} {\bibfnamefont
  {X.}~\bibnamefont {Hu}}, \bibinfo {author} {\bibfnamefont {H.-W.}\
  \bibnamefont {Jiang}}, \bibinfo {author} {\bibfnamefont {H.-O.}\ \bibnamefont
  {Li}}, \bibinfo {author} {\bibfnamefont {G.-C.}\ \bibnamefont {Guo}},\ and\
  \bibinfo {author} {\bibfnamefont {G.-P.}\ \bibnamefont {Guo}},\ }\bibfield
  {title} {\bibinfo {title} {Ultrafast coherent control of a hole spin qubit in
  a germanium quantum dot},\ }\href
  {https://doi.org/10.1038/s41467-021-27880-7} {\bibfield  {journal} {\bibinfo
  {journal} {Nature Communications}\ }\textbf {\bibinfo {volume} {13}},\
  \bibinfo {pages} {206} (\bibinfo {year} {2022})}\BibitemShut {NoStop}%
\bibitem [{\citenamefont {Maier}\ \emph {et~al.}(2013)\citenamefont {Maier},
  \citenamefont {Kloeffel},\ and\ \citenamefont {Loss}}]{PhysRevB.87.161305}%
  \BibitemOpen
  \bibfield  {author} {\bibinfo {author} {\bibfnamefont {F.}~\bibnamefont
  {Maier}}, \bibinfo {author} {\bibfnamefont {C.}~\bibnamefont {Kloeffel}},\
  and\ \bibinfo {author} {\bibfnamefont {D.}~\bibnamefont {Loss}},\ }\bibfield
  {title} {\bibinfo {title} {{Tunable $g$ factor and phonon-mediated hole spin
  relaxation in Ge/Si nanowire quantum dots}},\ }\href
  {https://doi.org/10.1103/PhysRevB.87.161305} {\bibfield  {journal} {\bibinfo
  {journal} {Phys. Rev. B}\ }\textbf {\bibinfo {volume} {87}},\ \bibinfo
  {pages} {161305} (\bibinfo {year} {2013})}\BibitemShut {NoStop}%
\bibitem [{\citenamefont {Adelsberger}\ \emph
  {et~al.}(2022{\natexlab{a}})\citenamefont {Adelsberger}, \citenamefont
  {Benito}, \citenamefont {Bosco}, \citenamefont {Klinovaja},\ and\
  \citenamefont {Loss}}]{PhysRevB.105.075308}%
  \BibitemOpen
  \bibfield  {author} {\bibinfo {author} {\bibfnamefont {C.}~\bibnamefont
  {Adelsberger}}, \bibinfo {author} {\bibfnamefont {M.}~\bibnamefont {Benito}},
  \bibinfo {author} {\bibfnamefont {S.}~\bibnamefont {Bosco}}, \bibinfo
  {author} {\bibfnamefont {J.}~\bibnamefont {Klinovaja}},\ and\ \bibinfo
  {author} {\bibfnamefont {D.}~\bibnamefont {Loss}},\ }\bibfield  {title}
  {\bibinfo {title} {Hole-spin qubits in {Ge} nanowire quantum dots: Interplay
  of orbital magnetic field, strain, and growth direction},\ }\href
  {https://doi.org/10.1103/PhysRevB.105.075308} {\bibfield  {journal} {\bibinfo
   {journal} {Phys. Rev. B}\ }\textbf {\bibinfo {volume} {105}},\ \bibinfo
  {pages} {075308} (\bibinfo {year} {2022}{\natexlab{a}})}\BibitemShut
  {NoStop}%
\bibitem [{\citenamefont {Adelsberger}\ \emph
  {et~al.}(2022{\natexlab{b}})\citenamefont {Adelsberger}, \citenamefont
  {Bosco}, \citenamefont {Klinovaja},\ and\ \citenamefont
  {Loss}}]{PhysRevB.106.235408}%
  \BibitemOpen
  \bibfield  {author} {\bibinfo {author} {\bibfnamefont {C.}~\bibnamefont
  {Adelsberger}}, \bibinfo {author} {\bibfnamefont {S.}~\bibnamefont {Bosco}},
  \bibinfo {author} {\bibfnamefont {J.}~\bibnamefont {Klinovaja}},\ and\
  \bibinfo {author} {\bibfnamefont {D.}~\bibnamefont {Loss}},\ }\bibfield
  {title} {\bibinfo {title} {Enhanced orbital magnetic field effects in {Ge}
  hole nanowires},\ }\href {https://doi.org/10.1103/PhysRevB.106.235408}
  {\bibfield  {journal} {\bibinfo  {journal} {Phys. Rev. B}\ }\textbf {\bibinfo
  {volume} {106}},\ \bibinfo {pages} {235408} (\bibinfo {year}
  {2022}{\natexlab{b}})}\BibitemShut {NoStop}%
\bibitem [{\citenamefont {Watzinger}\ \emph {et~al.}(2018)\citenamefont
  {Watzinger}, \citenamefont {Kuku{\v c}ka}, \citenamefont {Vuku{\v s}i{\'c}},
  \citenamefont {Gao}, \citenamefont {Wang}, \citenamefont {Sch{\"a}ffler},
  \citenamefont {Zhang},\ and\ \citenamefont {Katsaros}}]{Watzinger:2018aa}%
  \BibitemOpen
  \bibfield  {author} {\bibinfo {author} {\bibfnamefont {H.}~\bibnamefont
  {Watzinger}}, \bibinfo {author} {\bibfnamefont {J.}~\bibnamefont {Kuku{\v
  c}ka}}, \bibinfo {author} {\bibfnamefont {L.}~\bibnamefont {Vuku{\v
  s}i{\'c}}}, \bibinfo {author} {\bibfnamefont {F.}~\bibnamefont {Gao}},
  \bibinfo {author} {\bibfnamefont {T.}~\bibnamefont {Wang}}, \bibinfo {author}
  {\bibfnamefont {F.}~\bibnamefont {Sch{\"a}ffler}}, \bibinfo {author}
  {\bibfnamefont {J.-J.}\ \bibnamefont {Zhang}},\ and\ \bibinfo {author}
  {\bibfnamefont {G.}~\bibnamefont {Katsaros}},\ }\bibfield  {title} {\bibinfo
  {title} {A germanium hole spin qubit},\ }\href
  {https://doi.org/10.1038/s41467-018-06418-4} {\bibfield  {journal} {\bibinfo
  {journal} {Nature Communications}\ }\textbf {\bibinfo {volume} {9}},\
  \bibinfo {pages} {3902} (\bibinfo {year} {2018})}\BibitemShut {NoStop}%
\bibitem [{\citenamefont {Gao}\ \emph {et~al.}(2020)\citenamefont {Gao},
  \citenamefont {Wang}, \citenamefont {Watzinger}, \citenamefont {Hu},
  \citenamefont {Ran{\v c}i{\'c}}, \citenamefont {Zhang}, \citenamefont {Wang},
  \citenamefont {Yao}, \citenamefont {Wang}, \citenamefont {Kuku{\v c}ka},
  \citenamefont {Vuku{\v s}i{\'c}}, \citenamefont {Kloeffel}, \citenamefont
  {Loss}, \citenamefont {Liu}, \citenamefont {Katsaros},\ and\ \citenamefont
  {Zhang}}]{Gao2020AM}%
  \BibitemOpen
  \bibfield  {author} {\bibinfo {author} {\bibfnamefont {F.}~\bibnamefont
  {Gao}}, \bibinfo {author} {\bibfnamefont {J.-H.}\ \bibnamefont {Wang}},
  \bibinfo {author} {\bibfnamefont {H.}~\bibnamefont {Watzinger}}, \bibinfo
  {author} {\bibfnamefont {H.}~\bibnamefont {Hu}}, \bibinfo {author}
  {\bibfnamefont {M.~J.}\ \bibnamefont {Ran{\v c}i{\'c}}}, \bibinfo {author}
  {\bibfnamefont {J.-Y.}\ \bibnamefont {Zhang}}, \bibinfo {author}
  {\bibfnamefont {T.}~\bibnamefont {Wang}}, \bibinfo {author} {\bibfnamefont
  {Y.}~\bibnamefont {Yao}}, \bibinfo {author} {\bibfnamefont {G.-L.}\
  \bibnamefont {Wang}}, \bibinfo {author} {\bibfnamefont {J.}~\bibnamefont
  {Kuku{\v c}ka}}, \bibinfo {author} {\bibfnamefont {L.}~\bibnamefont {Vuku{\v
  s}i{\'c}}}, \bibinfo {author} {\bibfnamefont {C.}~\bibnamefont {Kloeffel}},
  \bibinfo {author} {\bibfnamefont {D.}~\bibnamefont {Loss}}, \bibinfo {author}
  {\bibfnamefont {F.}~\bibnamefont {Liu}}, \bibinfo {author} {\bibfnamefont
  {G.}~\bibnamefont {Katsaros}},\ and\ \bibinfo {author} {\bibfnamefont
  {J.-J.}\ \bibnamefont {Zhang}},\ }\bibfield  {title} {\bibinfo {title}
  {Site-controlled uniform {Ge/Si} hut wires with electrically tunable
  spin--orbit coupling},\ }\href
  {https://doi.org/https://doi.org/10.1002/adma.201906523} {\bibfield
  {journal} {\bibinfo  {journal} {Advanced Materials}\ }\textbf {\bibinfo
  {volume} {32}},\ \bibinfo {pages} {1906523} (\bibinfo {year}
  {2020})}\BibitemShut {NoStop}%
\bibitem [{\citenamefont {Higginbotham}\ \emph {et~al.}(2014)\citenamefont
  {Higginbotham}, \citenamefont {Kuemmeth}, \citenamefont {Larsen},
  \citenamefont {Fitzpatrick}, \citenamefont {Yao}, \citenamefont {Yan},
  \citenamefont {Lieber},\ and\ \citenamefont
  {Marcus}}]{PhysRevLett.112.216806}%
  \BibitemOpen
  \bibfield  {author} {\bibinfo {author} {\bibfnamefont {A.~P.}\ \bibnamefont
  {Higginbotham}}, \bibinfo {author} {\bibfnamefont {F.}~\bibnamefont
  {Kuemmeth}}, \bibinfo {author} {\bibfnamefont {T.~W.}\ \bibnamefont
  {Larsen}}, \bibinfo {author} {\bibfnamefont {M.}~\bibnamefont {Fitzpatrick}},
  \bibinfo {author} {\bibfnamefont {J.}~\bibnamefont {Yao}}, \bibinfo {author}
  {\bibfnamefont {H.}~\bibnamefont {Yan}}, \bibinfo {author} {\bibfnamefont
  {C.~M.}\ \bibnamefont {Lieber}},\ and\ \bibinfo {author} {\bibfnamefont
  {C.~M.}\ \bibnamefont {Marcus}},\ }\bibfield  {title} {\bibinfo {title}
  {Antilocalization of coulomb blockade in a {G}e/{S}i nanowire},\ }\href
  {https://doi.org/10.1103/PhysRevLett.112.216806} {\bibfield  {journal}
  {\bibinfo  {journal} {Phys. Rev. Lett.}\ }\textbf {\bibinfo {volume} {112}},\
  \bibinfo {pages} {216806} (\bibinfo {year} {2014})}\BibitemShut {NoStop}%
\bibitem [{\citenamefont {Nadj-Perge}\ \emph {et~al.}(2010)\citenamefont
  {Nadj-Perge}, \citenamefont {Frolov}, \citenamefont {Bakkers},\ and\
  \citenamefont {Kouwenhoven}}]{nadj2010spin}%
  \BibitemOpen
  \bibfield  {author} {\bibinfo {author} {\bibfnamefont {S.}~\bibnamefont
  {Nadj-Perge}}, \bibinfo {author} {\bibfnamefont {S.~M.}\ \bibnamefont
  {Frolov}}, \bibinfo {author} {\bibfnamefont {E.~P. A.~M.}\ \bibnamefont
  {Bakkers}},\ and\ \bibinfo {author} {\bibfnamefont {L.~P.}\ \bibnamefont
  {Kouwenhoven}},\ }\bibfield  {title} {\bibinfo {title} {Spin--orbit qubit in
  a semiconductor nanowire},\ }\href {https://doi.org/10.1038/nature09682}
  {\bibfield  {journal} {\bibinfo  {journal} {Nature}\ }\textbf {\bibinfo
  {volume} {468}},\ \bibinfo {pages} {1084} (\bibinfo {year}
  {2010})}\BibitemShut {NoStop}%
\bibitem [{\citenamefont {Trif}\ \emph {et~al.}(2008)\citenamefont {Trif},
  \citenamefont {Golovach},\ and\ \citenamefont {Loss}}]{trif2008spin}%
  \BibitemOpen
  \bibfield  {author} {\bibinfo {author} {\bibfnamefont {M.}~\bibnamefont
  {Trif}}, \bibinfo {author} {\bibfnamefont {V.~N.}\ \bibnamefont {Golovach}},\
  and\ \bibinfo {author} {\bibfnamefont {D.}~\bibnamefont {Loss}},\ }\bibfield
  {title} {\bibinfo {title} {{Spin dynamics in InAs nanowire quantum dots
  coupled to a transmission line}},\ }\href
  {https://doi.org/10.1103/PhysRevB.77.045434} {\bibfield  {journal} {\bibinfo
  {journal} {Phys. Rev. B}\ }\textbf {\bibinfo {volume} {77}},\ \bibinfo
  {pages} {045434} (\bibinfo {year} {2008})}\BibitemShut {NoStop}%
\bibitem [{\citenamefont {Kloeffel}\ \emph {et~al.}(2013)\citenamefont
  {Kloeffel}, \citenamefont {Trif}, \citenamefont {Stano},\ and\ \citenamefont
  {Loss}}]{PhysRevB.88.241405}%
  \BibitemOpen
  \bibfield  {author} {\bibinfo {author} {\bibfnamefont {C.}~\bibnamefont
  {Kloeffel}}, \bibinfo {author} {\bibfnamefont {M.}~\bibnamefont {Trif}},
  \bibinfo {author} {\bibfnamefont {P.}~\bibnamefont {Stano}},\ and\ \bibinfo
  {author} {\bibfnamefont {D.}~\bibnamefont {Loss}},\ }\bibfield  {title}
  {\bibinfo {title} {Circuit {QED} with hole-spin qubits in {G}e/{S}i nanowire
  quantum dots},\ }\href {https://doi.org/10.1103/PhysRevB.88.241405}
  {\bibfield  {journal} {\bibinfo  {journal} {Phys. Rev. B}\ }\textbf {\bibinfo
  {volume} {88}},\ \bibinfo {pages} {241405} (\bibinfo {year}
  {2013})}\BibitemShut {NoStop}%
\bibitem [{\citenamefont {Li}\ \emph {et~al.}(2013)\citenamefont {Li},
  \citenamefont {You}, \citenamefont {Sun},\ and\ \citenamefont
  {Nori}}]{RL2013}%
  \BibitemOpen
  \bibfield  {author} {\bibinfo {author} {\bibfnamefont {R.}~\bibnamefont
  {Li}}, \bibinfo {author} {\bibfnamefont {J.~Q.}\ \bibnamefont {You}},
  \bibinfo {author} {\bibfnamefont {C.~P.}\ \bibnamefont {Sun}},\ and\ \bibinfo
  {author} {\bibfnamefont {F.}~\bibnamefont {Nori}},\ }\bibfield  {title}
  {\bibinfo {title} {Controlling a nanowire spin-orbit qubit via
  electric-dipole spin resonance},\ }\href
  {https://doi.org/10.1103/PhysRevLett.111.086805} {\bibfield  {journal}
  {\bibinfo  {journal} {Phys. Rev. Lett.}\ }\textbf {\bibinfo {volume} {111}},\
  \bibinfo {pages} {086805} (\bibinfo {year} {2013})}\BibitemShut {NoStop}%
\bibitem [{\citenamefont {Li}(2018)}]{RL2018c}%
  \BibitemOpen
  \bibfield  {author} {\bibinfo {author} {\bibfnamefont {R.}~\bibnamefont
  {Li}},\ }\bibfield  {title} {\bibinfo {title} {A spin dephasing mechanism
  mediated by the interplay between the spin-orbit coupling and the
  asymmetrical confining potential in a semiconductor quantum dot},\ }\href
  {https://doi.org/10.1088/1361-648X/aadcb8} {\bibfield  {journal} {\bibinfo
  {journal} {Journal of Physics: Condensed Matter}\ }\textbf {\bibinfo {volume}
  {30}},\ \bibinfo {pages} {395304} (\bibinfo {year} {2018})}\BibitemShut
  {NoStop}%
\bibitem [{\citenamefont {Li}(2020)}]{RL2020}%
  \BibitemOpen
  \bibfield  {author} {\bibinfo {author} {\bibfnamefont {R.}~\bibnamefont
  {Li}},\ }\bibfield  {title} {\bibinfo {title} {Charge noise induced spin
  dephasing in a nanowire double quantum dot with spin{\textendash}orbit
  coupling},\ }\href {https://doi.org/10.1088/1361-648x/ab4933} {\bibfield
  {journal} {\bibinfo  {journal} {Journal of Physics: Condensed Matter}\
  }\textbf {\bibinfo {volume} {32}},\ \bibinfo {pages} {025305} (\bibinfo
  {year} {2020})}\BibitemShut {NoStop}%
\bibitem [{\citenamefont {Scappucci}\ \emph {et~al.}(2021)\citenamefont
  {Scappucci}, \citenamefont {Kloeffel}, \citenamefont {Zwanenburg},
  \citenamefont {Loss}, \citenamefont {Myronov}, \citenamefont {Zhang},
  \citenamefont {De~Franceschi}, \citenamefont {Katsaros},\ and\ \citenamefont
  {Veldhorst}}]{Scappucci:2021vk}%
  \BibitemOpen
  \bibfield  {author} {\bibinfo {author} {\bibfnamefont {G.}~\bibnamefont
  {Scappucci}}, \bibinfo {author} {\bibfnamefont {C.}~\bibnamefont {Kloeffel}},
  \bibinfo {author} {\bibfnamefont {F.~A.}\ \bibnamefont {Zwanenburg}},
  \bibinfo {author} {\bibfnamefont {D.}~\bibnamefont {Loss}}, \bibinfo {author}
  {\bibfnamefont {M.}~\bibnamefont {Myronov}}, \bibinfo {author} {\bibfnamefont
  {J.-J.}\ \bibnamefont {Zhang}}, \bibinfo {author} {\bibfnamefont
  {S.}~\bibnamefont {De~Franceschi}}, \bibinfo {author} {\bibfnamefont
  {G.}~\bibnamefont {Katsaros}},\ and\ \bibinfo {author} {\bibfnamefont
  {M.}~\bibnamefont {Veldhorst}},\ }\bibfield  {title} {\bibinfo {title} {The
  germanium quantum information route},\ }\href
  {https://doi.org/10.1038/s41578-020-00262-z} {\bibfield  {journal} {\bibinfo
  {journal} {Nature Reviews Materials}\ }\textbf {\bibinfo {volume} {6}},\
  \bibinfo {pages} {926} (\bibinfo {year} {2021})}\BibitemShut {NoStop}%
\bibitem [{\citenamefont {Khomitsky}\ and\ \citenamefont
  {Studenikin}(2022)}]{PhysRevB.106.195414}%
  \BibitemOpen
  \bibfield  {author} {\bibinfo {author} {\bibfnamefont {D.~V.}\ \bibnamefont
  {Khomitsky}}\ and\ \bibinfo {author} {\bibfnamefont {S.~A.}\ \bibnamefont
  {Studenikin}},\ }\bibfield  {title} {\bibinfo {title} {{Single-spin
  Landau-Zener-St\"uckelberg-Majorana interferometry of Zeeman-split states
  with strong spin-orbit interaction in a double quantum dot}},\ }\href
  {https://doi.org/10.1103/PhysRevB.106.195414} {\bibfield  {journal} {\bibinfo
   {journal} {Phys. Rev. B}\ }\textbf {\bibinfo {volume} {106}},\ \bibinfo
  {pages} {195414} (\bibinfo {year} {2022})}\BibitemShut {NoStop}%
\bibitem [{\citenamefont {Lutchyn}\ \emph {et~al.}(2010)\citenamefont
  {Lutchyn}, \citenamefont {Sau},\ and\ \citenamefont
  {Das~Sarma}}]{PhysRevLett.105.077001}%
  \BibitemOpen
  \bibfield  {author} {\bibinfo {author} {\bibfnamefont {R.~M.}\ \bibnamefont
  {Lutchyn}}, \bibinfo {author} {\bibfnamefont {J.~D.}\ \bibnamefont {Sau}},\
  and\ \bibinfo {author} {\bibfnamefont {S.}~\bibnamefont {Das~Sarma}},\
  }\bibfield  {title} {\bibinfo {title} {Majorana fermions and a topological
  phase transition in semiconductor-superconductor heterostructures},\ }\href
  {https://doi.org/10.1103/PhysRevLett.105.077001} {\bibfield  {journal}
  {\bibinfo  {journal} {Phys. Rev. Lett.}\ }\textbf {\bibinfo {volume} {105}},\
  \bibinfo {pages} {077001} (\bibinfo {year} {2010})}\BibitemShut {NoStop}%
\bibitem [{\citenamefont {Oreg}\ \emph {et~al.}(2010)\citenamefont {Oreg},
  \citenamefont {Refael},\ and\ \citenamefont {von
  Oppen}}]{PhysRevLett.105.177002}%
  \BibitemOpen
  \bibfield  {author} {\bibinfo {author} {\bibfnamefont {Y.}~\bibnamefont
  {Oreg}}, \bibinfo {author} {\bibfnamefont {G.}~\bibnamefont {Refael}},\ and\
  \bibinfo {author} {\bibfnamefont {F.}~\bibnamefont {von Oppen}},\ }\bibfield
  {title} {\bibinfo {title} {Helical liquids and majorana bound states in
  quantum wires},\ }\href {https://doi.org/10.1103/PhysRevLett.105.177002}
  {\bibfield  {journal} {\bibinfo  {journal} {Phys. Rev. Lett.}\ }\textbf
  {\bibinfo {volume} {105}},\ \bibinfo {pages} {177002} (\bibinfo {year}
  {2010})}\BibitemShut {NoStop}%
\bibitem [{\citenamefont {Maier}\ \emph {et~al.}(2014)\citenamefont {Maier},
  \citenamefont {Klinovaja},\ and\ \citenamefont {Loss}}]{PhysRevB.90.195421}%
  \BibitemOpen
  \bibfield  {author} {\bibinfo {author} {\bibfnamefont {F.}~\bibnamefont
  {Maier}}, \bibinfo {author} {\bibfnamefont {J.}~\bibnamefont {Klinovaja}},\
  and\ \bibinfo {author} {\bibfnamefont {D.}~\bibnamefont {Loss}},\ }\bibfield
  {title} {\bibinfo {title} {Majorana fermions in {Ge/Si} hole nanowires},\
  }\href {https://doi.org/10.1103/PhysRevB.90.195421} {\bibfield  {journal}
  {\bibinfo  {journal} {Phys. Rev. B}\ }\textbf {\bibinfo {volume} {90}},\
  \bibinfo {pages} {195421} (\bibinfo {year} {2014})}\BibitemShut {NoStop}%
\bibitem [{\citenamefont {Kloeffel}\ \emph {et~al.}(2011)\citenamefont
  {Kloeffel}, \citenamefont {Trif},\ and\ \citenamefont
  {Loss}}]{PhysRevB.84.195314}%
  \BibitemOpen
  \bibfield  {author} {\bibinfo {author} {\bibfnamefont {C.}~\bibnamefont
  {Kloeffel}}, \bibinfo {author} {\bibfnamefont {M.}~\bibnamefont {Trif}},\
  and\ \bibinfo {author} {\bibfnamefont {D.}~\bibnamefont {Loss}},\ }\bibfield
  {title} {\bibinfo {title} {Strong spin-orbit interaction and helical hole
  states in {Ge/Si} nanowires},\ }\href
  {https://doi.org/10.1103/PhysRevB.84.195314} {\bibfield  {journal} {\bibinfo
  {journal} {Phys. Rev. B}\ }\textbf {\bibinfo {volume} {84}},\ \bibinfo
  {pages} {195314} (\bibinfo {year} {2011})}\BibitemShut {NoStop}%
\bibitem [{\citenamefont {Li}(2021)}]{RL2021}%
  \BibitemOpen
  \bibfield  {author} {\bibinfo {author} {\bibfnamefont {R.}~\bibnamefont
  {Li}},\ }\bibfield  {title} {\bibinfo {title} {Low-energy subband
  wave-functions and effective g-factor of one-dimensional hole gas},\ }\href
  {https://doi.org/10.1088/1361-648x/ac0d18} {\bibfield  {journal} {\bibinfo
  {journal} {Journal of Physics: Condensed Matter}\ }\textbf {\bibinfo {volume}
  {33}},\ \bibinfo {pages} {355302} (\bibinfo {year} {2021})}\BibitemShut
  {NoStop}%
\bibitem [{\citenamefont {Li}(2022)}]{RL2022a}%
  \BibitemOpen
  \bibfield  {author} {\bibinfo {author} {\bibfnamefont {R.}~\bibnamefont
  {Li}},\ }\bibfield  {title} {\bibinfo {title} {Searching strong `spin'-orbit
  coupled one-dimensional hole gas in strong magnetic fields},\ }\href
  {https://doi.org/10.1088/1361-648x/ac37da} {\bibfield  {journal} {\bibinfo
  {journal} {Journal of Physics: Condensed Matter}\ }\textbf {\bibinfo {volume}
  {34}},\ \bibinfo {pages} {075301} (\bibinfo {year} {2022})}\BibitemShut
  {NoStop}%
\bibitem [{\citenamefont {Kloeffel}\ \emph {et~al.}(2018)\citenamefont
  {Kloeffel}, \citenamefont {Ran\ifmmode \check{c}\else
  \v{c}\fi{}i\ifmmode~\acute{c}\else \'{c}\fi{}},\ and\ \citenamefont
  {Loss}}]{PhysRevB.97.235422}%
  \BibitemOpen
  \bibfield  {author} {\bibinfo {author} {\bibfnamefont {C.}~\bibnamefont
  {Kloeffel}}, \bibinfo {author} {\bibfnamefont {M.~J.}\ \bibnamefont
  {Ran\ifmmode \check{c}\else \v{c}\fi{}i\ifmmode~\acute{c}\else \'{c}\fi{}}},\
  and\ \bibinfo {author} {\bibfnamefont {D.}~\bibnamefont {Loss}},\ }\bibfield
  {title} {\bibinfo {title} {Direct rashba spin-orbit interaction in {Si} and
  {Ge} nanowires with different growth directions},\ }\href
  {https://doi.org/10.1103/PhysRevB.97.235422} {\bibfield  {journal} {\bibinfo
  {journal} {Phys. Rev. B}\ }\textbf {\bibinfo {volume} {97}},\ \bibinfo
  {pages} {235422} (\bibinfo {year} {2018})}\BibitemShut {NoStop}%
\bibitem [{\citenamefont {Luo}\ \emph {et~al.}(2017)\citenamefont {Luo},
  \citenamefont {Li},\ and\ \citenamefont {Zunger}}]{PhysRevLett.119.126401}%
  \BibitemOpen
  \bibfield  {author} {\bibinfo {author} {\bibfnamefont {J.-W.}\ \bibnamefont
  {Luo}}, \bibinfo {author} {\bibfnamefont {S.-S.}\ \bibnamefont {Li}},\ and\
  \bibinfo {author} {\bibfnamefont {A.}~\bibnamefont {Zunger}},\ }\bibfield
  {title} {\bibinfo {title} {Rapid transition of the hole rashba effect from
  strong field dependence to saturation in semiconductor nanowires},\ }\href
  {https://doi.org/10.1103/PhysRevLett.119.126401} {\bibfield  {journal}
  {\bibinfo  {journal} {Phys. Rev. Lett.}\ }\textbf {\bibinfo {volume} {119}},\
  \bibinfo {pages} {126401} (\bibinfo {year} {2017})}\BibitemShut {NoStop}%
\bibitem [{\citenamefont {Li}\ and\ \citenamefont {Zhang}(2023)}]{RL2023a}%
  \BibitemOpen
  \bibfield  {author} {\bibinfo {author} {\bibfnamefont {R.}~\bibnamefont
  {Li}}\ and\ \bibinfo {author} {\bibfnamefont {H.}~\bibnamefont {Zhang}},\
  }\bibfield  {title} {\bibinfo {title} {Electrical manipulation of a hole
  `spin'--orbit qubit in nanowire quantum dot: The nontrivial magnetic field
  effects},\ }\bibfield  {booktitle} {\emph {\bibinfo {booktitle} {Chinese
  Physics B}},\ }\href {https://doi.org/10.1088/1674-1056/ac873b} {\ \textbf
  {\bibinfo {volume} {32}},\ \bibinfo {pages} {030308} (\bibinfo {year}
  {2023})}\BibitemShut {NoStop}%
\bibitem [{\citenamefont {Li}\ and\ \citenamefont {Qi}(2023)}]{RL2023b}%
  \BibitemOpen
  \bibfield  {author} {\bibinfo {author} {\bibfnamefont {R.}~\bibnamefont
  {Li}}\ and\ \bibinfo {author} {\bibfnamefont {X.-Y.}\ \bibnamefont {Qi}},\
  }\bibfield  {title} {\bibinfo {title} {Two-band description of the strong
  `spin'-orbit coupled one-dimensional hole gas in a cylindrical ge nanowire},\
  }\bibfield  {booktitle} {\emph {\bibinfo {booktitle} {Journal of Physics:
  Condensed Matter}},\ }\href {https://doi.org/10.1088/1361-648X/acb8f5} {\
  \textbf {\bibinfo {volume} {35}},\ \bibinfo {pages} {135302} (\bibinfo {year}
  {2023})}\BibitemShut {NoStop}%
\bibitem [{\citenamefont {Csontos}\ \emph {et~al.}(2009)\citenamefont
  {Csontos}, \citenamefont {Brusheim}, \citenamefont {Z\"ulicke},\ and\
  \citenamefont {Xu}}]{PhysRevB.79.155323}%
  \BibitemOpen
  \bibfield  {author} {\bibinfo {author} {\bibfnamefont {D.}~\bibnamefont
  {Csontos}}, \bibinfo {author} {\bibfnamefont {P.}~\bibnamefont {Brusheim}},
  \bibinfo {author} {\bibfnamefont {U.}~\bibnamefont {Z\"ulicke}},\ and\
  \bibinfo {author} {\bibfnamefont {H.~Q.}\ \bibnamefont {Xu}},\ }\bibfield
  {title} {\bibinfo {title} {Spin-$\frac{3}{2}$ physics of semiconductor hole
  nanowires: Valence-band mixing and tunable interplay between bulk-material
  and orbital bound-state spin splittings},\ }\href
  {https://doi.org/10.1103/PhysRevB.79.155323} {\bibfield  {journal} {\bibinfo
  {journal} {Phys. Rev. B}\ }\textbf {\bibinfo {volume} {79}},\ \bibinfo
  {pages} {155323} (\bibinfo {year} {2009})}\BibitemShut {NoStop}%
\bibitem [{\citenamefont {Lawaetz}(1971)}]{PhysRevB.4.3460}%
  \BibitemOpen
  \bibfield  {author} {\bibinfo {author} {\bibfnamefont {P.}~\bibnamefont
  {Lawaetz}},\ }\bibfield  {title} {\bibinfo {title} {Valence-band parameters
  in cubic semiconductors},\ }\href {https://doi.org/10.1103/PhysRevB.4.3460}
  {\bibfield  {journal} {\bibinfo  {journal} {Phys. Rev. B}\ }\textbf {\bibinfo
  {volume} {4}},\ \bibinfo {pages} {3460} (\bibinfo {year} {1971})}\BibitemShut
  {NoStop}%
\bibitem [{\citenamefont {Budkin}\ and\ \citenamefont
  {Tarasenko}(2022)}]{PhysRevB.105.L161301}%
  \BibitemOpen
  \bibfield  {author} {\bibinfo {author} {\bibfnamefont {G.~V.}\ \bibnamefont
  {Budkin}}\ and\ \bibinfo {author} {\bibfnamefont {S.~A.}\ \bibnamefont
  {Tarasenko}},\ }\bibfield  {title} {\bibinfo {title} {Spin splitting in
  low-symmetry quantum wells beyond {Rashba} and {Dresselhaus} terms},\ }\href
  {https://doi.org/10.1103/PhysRevB.105.L161301} {\bibfield  {journal}
  {\bibinfo  {journal} {Phys. Rev. B}\ }\textbf {\bibinfo {volume} {105}},\
  \bibinfo {pages} {L161301} (\bibinfo {year} {2022})}\BibitemShut {NoStop}%
\bibitem [{\citenamefont {Roddaro}\ \emph {et~al.}(2008)\citenamefont
  {Roddaro}, \citenamefont {Fuhrer}, \citenamefont {Brusheim}, \citenamefont
  {Fasth}, \citenamefont {Xu}, \citenamefont {Samuelson}, \citenamefont
  {Xiang},\ and\ \citenamefont {Lieber}}]{PhysRevLett.101.186802}%
  \BibitemOpen
  \bibfield  {author} {\bibinfo {author} {\bibfnamefont {S.}~\bibnamefont
  {Roddaro}}, \bibinfo {author} {\bibfnamefont {A.}~\bibnamefont {Fuhrer}},
  \bibinfo {author} {\bibfnamefont {P.}~\bibnamefont {Brusheim}}, \bibinfo
  {author} {\bibfnamefont {C.}~\bibnamefont {Fasth}}, \bibinfo {author}
  {\bibfnamefont {H.~Q.}\ \bibnamefont {Xu}}, \bibinfo {author} {\bibfnamefont
  {L.}~\bibnamefont {Samuelson}}, \bibinfo {author} {\bibfnamefont
  {J.}~\bibnamefont {Xiang}},\ and\ \bibinfo {author} {\bibfnamefont {C.~M.}\
  \bibnamefont {Lieber}},\ }\bibfield  {title} {\bibinfo {title} {Spin states
  of holes in $\mathrm{Ge}/\mathrm{Si}$ nanowire quantum dots},\ }\href
  {https://doi.org/10.1103/PhysRevLett.101.186802} {\bibfield  {journal}
  {\bibinfo  {journal} {Phys. Rev. Lett.}\ }\textbf {\bibinfo {volume} {101}},\
  \bibinfo {pages} {186802} (\bibinfo {year} {2008})}\BibitemShut {NoStop}%
\bibitem [{\citenamefont {Landau}\ and\ \citenamefont
  {Lifshitz}(1965)}]{landau1965quantum}%
  \BibitemOpen
  \bibfield  {author} {\bibinfo {author} {\bibfnamefont {L.~D.}\ \bibnamefont
  {Landau}}\ and\ \bibinfo {author} {\bibfnamefont {E.~M.}\ \bibnamefont
  {Lifshitz}},\ }\href@noop {} {\emph {\bibinfo {title} {Quantum Mechanics}}}\
  (\bibinfo  {publisher} {Pergamon, New York},\ \bibinfo {year}
  {1965})\BibitemShut {NoStop}%
\bibitem [{\citenamefont {Sercel}\ and\ \citenamefont
  {Vahala}(1990)}]{PhysRevB.42.3690}%
  \BibitemOpen
  \bibfield  {author} {\bibinfo {author} {\bibfnamefont {P.~C.}\ \bibnamefont
  {Sercel}}\ and\ \bibinfo {author} {\bibfnamefont {K.~J.}\ \bibnamefont
  {Vahala}},\ }\bibfield  {title} {\bibinfo {title} {Analytical formalism for
  determining quantum-wire and quantum-dot band structure in the multiband
  envelope-function approximation},\ }\href
  {https://doi.org/10.1103/PhysRevB.42.3690} {\bibfield  {journal} {\bibinfo
  {journal} {Phys. Rev. B}\ }\textbf {\bibinfo {volume} {42}},\ \bibinfo
  {pages} {3690} (\bibinfo {year} {1990})}\BibitemShut {NoStop}%
\bibitem [{\citenamefont {Sweeny}\ \emph {et~al.}(1988)\citenamefont {Sweeny},
  \citenamefont {Xu},\ and\ \citenamefont {Shur}}]{sweeny1988hole}%
  \BibitemOpen
  \bibfield  {author} {\bibinfo {author} {\bibfnamefont {M.}~\bibnamefont
  {Sweeny}}, \bibinfo {author} {\bibfnamefont {J.}~\bibnamefont {Xu}},\ and\
  \bibinfo {author} {\bibfnamefont {M.}~\bibnamefont {Shur}},\ }\bibfield
  {title} {\bibinfo {title} {Hole subbands in one-dimensional quantum well
  wires},\ }\href {https://doi.org/10.1016/0749-6036(88)90249-2} {\bibfield
  {journal} {\bibinfo  {journal} {Superlattices and Microstructures}\ }\textbf
  {\bibinfo {volume} {4}},\ \bibinfo {pages} {623} (\bibinfo {year}
  {1988})}\BibitemShut {NoStop}%
\bibitem [{\citenamefont {Sun}\ \emph {et~al.}(2009)\citenamefont {Sun},
  \citenamefont {Thompson},\ and\ \citenamefont {Nishida}}]{sun2009strain}%
  \BibitemOpen
  \bibfield  {author} {\bibinfo {author} {\bibfnamefont {Y.}~\bibnamefont
  {Sun}}, \bibinfo {author} {\bibfnamefont {S.~E.}\ \bibnamefont {Thompson}},\
  and\ \bibinfo {author} {\bibfnamefont {T.}~\bibnamefont {Nishida}},\
  }\href@noop {} {\emph {\bibinfo {title} {Strain effect in semiconductors:
  theory and device applications}}}\ (\bibinfo  {publisher} {Springer Science
  \& Business Media, New York},\ \bibinfo {year} {2009})\BibitemShut {NoStop}%
\bibitem [{\citenamefont {Bir}\ \emph {et~al.}(1974)\citenamefont {Bir},
  \citenamefont {Pikus},\ and\ \citenamefont {Louvish}}]{bir1974symmetry}%
  \BibitemOpen
  \bibfield  {author} {\bibinfo {author} {\bibfnamefont {G.~L.}\ \bibnamefont
  {Bir}}, \bibinfo {author} {\bibfnamefont {G.~E.}\ \bibnamefont {Pikus}},\
  and\ \bibinfo {author} {\bibfnamefont {D.}~\bibnamefont {Louvish}},\
  }\href@noop {} {\emph {\bibinfo {title} {Symmetry and strain-induced effects
  in semiconductors}}},\ Vol.\ \bibinfo {volume} {484}\ (\bibinfo  {publisher}
  {Wiley New York},\ \bibinfo {year} {1974})\BibitemShut {NoStop}%
\bibitem [{\citenamefont {Kloeffel}\ \emph {et~al.}(2014)\citenamefont
  {Kloeffel}, \citenamefont {Trif},\ and\ \citenamefont
  {Loss}}]{PhysRevB.90.115419}%
  \BibitemOpen
  \bibfield  {author} {\bibinfo {author} {\bibfnamefont {C.}~\bibnamefont
  {Kloeffel}}, \bibinfo {author} {\bibfnamefont {M.}~\bibnamefont {Trif}},\
  and\ \bibinfo {author} {\bibfnamefont {D.}~\bibnamefont {Loss}},\ }\bibfield
  {title} {\bibinfo {title} {Acoustic phonons and strain in core/shell
  nanowires},\ }\href {https://doi.org/10.1103/PhysRevB.90.115419} {\bibfield
  {journal} {\bibinfo  {journal} {Phys. Rev. B}\ }\textbf {\bibinfo {volume}
  {90}},\ \bibinfo {pages} {115419} (\bibinfo {year} {2014})}\BibitemShut
  {NoStop}%
\bibitem [{\citenamefont {Liao}\ \emph {et~al.}(2015)\citenamefont {Liao},
  \citenamefont {Luo}, \citenamefont {Yang}, \citenamefont {Chen},\ and\
  \citenamefont {Xu}}]{doi:10.1063/1.4929412}%
  \BibitemOpen
  \bibfield  {author} {\bibinfo {author} {\bibfnamefont {G.}~\bibnamefont
  {Liao}}, \bibinfo {author} {\bibfnamefont {N.}~\bibnamefont {Luo}}, \bibinfo
  {author} {\bibfnamefont {Z.}~\bibnamefont {Yang}}, \bibinfo {author}
  {\bibfnamefont {K.}~\bibnamefont {Chen}},\ and\ \bibinfo {author}
  {\bibfnamefont {H.~Q.}\ \bibnamefont {Xu}},\ }\bibfield  {title} {\bibinfo
  {title} {{Electronic structures of [001]- and [111]-oriented InSb and GaSb
  free-standing nanowires}},\ }\href {https://doi.org/10.1063/1.4929412}
  {\bibfield  {journal} {\bibinfo  {journal} {Journal of Applied Physics}\
  }\textbf {\bibinfo {volume} {118}},\ \bibinfo {pages} {094308} (\bibinfo
  {year} {2015})}\BibitemShut {NoStop}%
\bibitem [{\citenamefont {Luo}\ \emph {et~al.}(2016)\citenamefont {Luo},
  \citenamefont {Liao},\ and\ \citenamefont {Xu}}]{doi:10.1063/1.4972987}%
  \BibitemOpen
  \bibfield  {author} {\bibinfo {author} {\bibfnamefont {N.}~\bibnamefont
  {Luo}}, \bibinfo {author} {\bibfnamefont {G.}~\bibnamefont {Liao}},\ and\
  \bibinfo {author} {\bibfnamefont {H.~Q.}\ \bibnamefont {Xu}},\ }\bibfield
  {title} {\bibinfo {title} {k.p theory of freestanding narrow band gap
  semiconductor nanowires},\ }\href {https://doi.org/10.1063/1.4972987}
  {\bibfield  {journal} {\bibinfo  {journal} {AIP Advances}\ }\textbf {\bibinfo
  {volume} {6}},\ \bibinfo {pages} {125109} (\bibinfo {year}
  {2016})}\BibitemShut {NoStop}%
\bibitem [{\citenamefont {Kash}\ \emph {et~al.}(1994)\citenamefont {Kash},
  \citenamefont {Zachau}, \citenamefont {Tischler},\ and\ \citenamefont
  {Ekenberg}}]{KASH1994251}%
  \BibitemOpen
  \bibfield  {author} {\bibinfo {author} {\bibfnamefont {J.}~\bibnamefont
  {Kash}}, \bibinfo {author} {\bibfnamefont {M.}~\bibnamefont {Zachau}},
  \bibinfo {author} {\bibfnamefont {M.}~\bibnamefont {Tischler}},\ and\
  \bibinfo {author} {\bibfnamefont {U.}~\bibnamefont {Ekenberg}},\ }\bibfield
  {title} {\bibinfo {title} {Optical measurements of warped valence bands in
  quantum wells},\ }\href
  {https://doi.org/https://doi.org/10.1016/0039-6028(94)90895-8} {\bibfield
  {journal} {\bibinfo  {journal} {Surface Science}\ }\textbf {\bibinfo {volume}
  {305}},\ \bibinfo {pages} {251} (\bibinfo {year} {1994})}\BibitemShut
  {NoStop}%
\end{thebibliography}%
\end{document}